\documentclass[12pt,oneside,onecolumn]{elsarticle}

\usepackage[ansinew]{inputenc}
\usepackage[english]{babel}
\usepackage{ifthen,shortvrb}
\usepackage[fleqn]{amsmath}
\usepackage{amsxtra,amsfonts,latexsym,amssymb,euscript}
\usepackage{amstext}
\usepackage{graphicx}
\usepackage{dcolumn}
\usepackage{subfigure}
\usepackage{endfloat}

\newcommand{\db}[2]{\frac{\partial #1}{\partial #2}}

\newcommand{\ssa}{\rule{5mm}{0mm}}
\newcommand{\ssb}[1]{\ssa\mbox{#1}\ssa}

\newcommand{\Estar}{E^{\raisebox{0.5mm}{$ *$}}}

\begin{document}

\begin{frontmatter}
\journal{Journal of the Mechanics and Physics of Solids}
\title{JKR solution for an anisotropic half space}
\author[ME]{J. R. Barber\corref{JRB}}
\author[B]{M.Ciavarella}
\address[ME]{Department of Mechanical Engineering,
University of Michigan, Ann Arbor, MI 48109-2125, U.S.A.}
\address[B]{CEMEC-Politecnico di Bari, Bari - Italy.}
\cortext[JRB]{jbarber@umich.edu.}

\begin{abstract}
In this paper, the classical JKR theory of the adhesive contact of isotropic elastic spheres is extended to consider the effect of anisotropic elasticity. The contact area will then generally be non-circular, but in many cases it can reasonably be approximated by an ellipse whose dimensions are determined by imposing the energy release rate criterion at the ends of the major and minor axes. Analytical expressions are obtained for the relations between the contact force, the normal displacement and the ellipse semi-axes. It is found that the eccentricity of the contact area decreases during tensile loading and for cases when the point load solution can be accurately described by only one Fourier term, it is almost circular at pull-off, permitting an exact closed form solution for this case.  As in the isotropic JKR solution, the pull-off force is independent of the mean elastic modulus, but we find that anisotropy increases the pull-off force and this effect can be quite significant.

\end{abstract}
\end{frontmatter}

\section{Introduction}

The JKR theory of contact between isotropic elastic spheres including interatomic adhesion \cite{JKR} is very widely used (3500 citations and counting), primarily because it  provides relatively simple theoretical predictions of the effect of adhesive forces in contact situations. In particular, the force needed to separate the bodies (the `pull-off force') is given by
\begin{equation}
F=\frac{3\pi R\Delta\gamma}{2}\;, \label{JKRF}
\end{equation}
where $R$ is the composite radius of the contacting spheres and $\Delta\gamma$ is the interface energy.

The theory applies strictly in the limit where the Tabor parameter 
\begin{equation*}
\mu=\left(\frac{R(\Delta\gamma)^2}{\Estar\,\!^2\epsilon^3}
\right)^{1/3}\gg 1\;, 
\end{equation*}
where $\Estar$ is the composite elastic modulus, and $\epsilon$ is a dimension characterizing the length over which the interatomic forces are significant. However, numerical treatments of the more general problem \cite{MYD,Greenwood-adhesion} show that the pull-off force varies rather modestly with $\mu$ and indeed in the opposite limit where $\mu\rightarrow0$, we recover the Bradley solution which exceeds the JKR prediction only by a factor of 4/3, and which has the same parametric dependence. It is remarkable that the pull-off force is independent of the modulus $\Estar$. This can be shown to be a consequence of the self-similar nature of the contact problem and the quadratic shape of the surfaces \cite{JRB-adhesion}. However, for the corresponding two-dimensional problem of a cylinder contacting a plane, the pull-off force varies with the 1/3rd power of $\Estar$ \cite{Chaudhury}.

Most of the applications and extensions of the JKR theory involve contact problems at very small length scales, since this is the range in which interatomic adhesive forces are most significant.  The theory is based on the assumption that the contacting bodies be capable of approximation by half spaces of linear elastic materials, which is certainly an oversimplification in most biological and animal locomotion applications, but these assumptions are more reasonable for microindentation or AFM contacts with elastic materials at light loads, and indeed such experiments are often used to estimate the elastic properties of such materials at small length scales \cite{Pharr1,Pharr2,Vlassak}.

Most materials exhibit significant anisotropy at the microscale, either because of crystalline structure or because the material has some more complex structural composition at the nanoscale. If the anisotropy is relatively mild, we might expect to get a reasonable prediction of the indentation behaviour by using the
original JKR solution, with an appropriate `mean' elastic modulus. However, when the materials are anisotropic, we anticipate that the contact area will cease to be circular and this might be expected to infuence the load-displacement relation significantly. In particular, noting that the pull-off force is independent of the elastic modulus for the isotropic case, we might ask whether this will be influenced by the degree of anisotropy.  These are the questions that we shall investigate in the present paper.

\section{Normal loading of the anistropic elastic half space}

If a concentrated normal compressive force $F$ is applied at the origin to the surface of the half space $z>0$, self-similarity and equilibrium considerations dictate that the normal surface displacement take the form \cite{Willis-Flat}
\begin{equation}
u(r,\theta)\equiv u_z(r,\theta,0)=\frac{Fh(\theta)}{r} \label{G3}
\end{equation}
in cylindrical polar coordinates $(r,\theta,z)$. Also, the reciprocal theorem demands that 
\begin{equation}
u(r,\theta+\pi)=u(r,\theta)
\end{equation}
and hence the function $h(\theta)$ must be capable of Fourier expansion in the form 
\begin{equation}
h(\theta)=h_0\left[1+\sum_{m=1}^\infty a_m\cos(2m\theta)+b_m\sin(2m\theta)\right]\;,  \label{htheta}
\end{equation}
\cite{VlassakJRB} where we have extracted the dimensional `mean' compliance $h_0$, so that the remaining coefficients $a_m, b_m$ are dimensionless measures of the degree of anisotropy.

If the three-dimensional Green's function is defined by equation (\ref{G3}), the corresponding two-dimensional (plane strain) result can be obtained by defining a uniform distribution of forces $F$ per unit length along an appropriate line. For example, if the distribution is imposed along the infinite line $x=0, -\infty<y<\infty$, the resulting value of $\mbox{$\partial$} u/\mbox{$\partial$} x$ at the point $(x,0)$ is obtained as 
\begin{equation}
\db{u}{x}=\int_{-\infty}^\infty\left(\db{u}{r}\cos\theta-\frac{1}{r}\db{u}{\theta}
\sin\theta\right)dy\;.
\end{equation}
Susbtituting for $u$ from (\ref{G3}), writing 
\begin{equation}
y=-x\tan\theta\;;\;\;\;dy=-\frac{x d\theta}{\cos^2\theta}\;;\;\;\;r=\frac{x}{\cos\theta}\;,
\end{equation}
and evaluating the resulting integral, we obtain 
\begin{equation}
\db{u}{x}=-\frac{2Fh(\pi/2)}{x}
\end{equation}
for the plane strain Green's function appropriate to fields that are independent of $y$. 

Since the Cartesian coordinate system can be chosen arbitrarily, we conclude that the function $h(\theta)$ in equation (\ref{G3}) is proportional to the plane strain compliance in the direction perpendicular to $\theta$, and this can be obtained by applying the Stroh formalism to the general anisotropic constants $c_{ijkl}$ rotated through $\theta+\pi/2$ using the tensor transformation rules \cite{Vlassak,Gao-Pharr}.

\subsection{Approximate results for orthotropic and transversely isotropic materials}

Since the principal effect of anisotropy is to change the eccentricity of the contact area, it seems likely that the deviation from axisymmetry will be dominated by the $\cos(2\theta)$ term in equation (\ref{htheta}). If the material is orthotropic, a simple approximation to the function $h(\theta)$ can then be obtained as
\begin{equation}
h(\theta)=\frac{1}{2}\left[h(0)+h\left(\frac{\pi}{2}\right)\right]+\frac{1}{2}\left[h(0)-h\left(\frac{\pi}{2}\right)\right]\cos(2\theta)\;. \label{DU1}
\end{equation}
Delafargue \& Ulm \cite{Delafargue} show that this gives a good approximation to the more exact result for examples of orthotropic and transversely isotropic materials when the surface is a plane of symmetry. 

For the orthotropic case, if we take the surface to be defined by $x_1=0$ and measure $\theta$ from the $x_2$-axis, $h(0)$ and $h(\pi/2)$ are given by
\begin{eqnarray}
h(0)&=&\frac{1}{2\pi }\sqrt{\frac{C_{22}}{C_{11}C_{22}-C_{12}^{2}}\left( 
\frac{1}{C_{66}}+\frac{2}{C_{12}+\sqrt{C_{11}C_{22}}}\right) } \label{O1}\\
h\left(\frac{\pi}{2}\right)&=&\frac{1}{2\pi }\sqrt{\frac{C_{33}}{C_{11}C_{33}-C_{13}^{2}}\left( 
\frac{1}{C_{55}}+\frac{2}{C_{13}+\sqrt{C_{11}C_{33}}}\right) }\;,\label{O2}
\end{eqnarray}
 where we use the usual reduced notation $11\!\rightarrow\!1,\;22\!\rightarrow\!2,\;33\!\rightarrow\!3,\;23\!\rightarrow\!4,\;31\!\rightarrow\!5,\;12\!\rightarrow\!6$.

Equations (\ref{O1},\ref{O2}) apply also in the special case of transverse isotropy with appropriate values for the constants. For example, if the material is isotropic in the $x_1x_2$-plane, $C_{22}=C_{11}, 2C_{66}=C_{11}-C_{12}$ and $h(\pi/2)$ remains unchanged, but $h(0)$ reduces to
\[
h(0)=\frac{C_{11}}{\pi(C_{11}^{2}-C_{12}^{2})}\;,
\]
which is identical with the indentation modulus of an isotropic material with elastic constants $C_{11}, C_{12}$ \cite{Delafargue}.

If the Green's function (\ref{htheta}) is approximated in the form (\ref{DU1}), the only non-zero coefficient is
\begin{equation}
a_1=\frac{h(\pi/2)-h(0)}{h(\pi/2)+h(0)}\;. \label{a1-approx}
\end{equation}

Table 1 gives elastic moduli (from Freund \& Suresh \cite{Freund}) and the resulting dimensionless parameter $a_1$ for a few hexagonal crystals, which exhibit transverse isotropic behaviour. We consider the case where the surface is orthogonal to the plane of isotropy, so that the direction of indentation lies in this plane and the directional compliance modulus $h(\theta)$ is not axisymmetric.

\begin{center}
\begin{tabular}{|l|c|c|c|c|c|c|c|c|}
\hline
&$C_{11}$&$C_{33}$&$C_{44}$&$C_{12}$&$C_{13}$&$h(\pi/2)$&$h(0)$&$a_1$\\
\hline
&(GPa)&(GPa)&(GPa)&(GPa)&(GPa)&(MPa)$^{-1}$&(MPa)$^{-1}$&\\
\hline
\hline
cadmium&115.8&	51.4&	20.4&	39.8&	40.6&	28.08&	19.58&	$-$0.178\\
\hline
cobalt&307&	358.1&	78.3&	165	&103&	\;7.91&	\;9.16&	\;\;\;0.073\\
\hline
graphite&1160&46.6&2.3&290&109&22.07&\;1.84&$-$0.846\\
\hline
magnesium&59.7&	61.7&	16.4&	26.2&	21.7&	40.46&	41.49&	\;\;\;0.013\\
\hline
zinc&161&61&38.3&34.2&50.1&18.16&13.01&	$-$0.165\\
\hline
titanium&162.4&	180.7&	46.7&	92	&69&	14.78&	18.14&	\;\;\;0.102\\
\hline
\end{tabular}

\vspace{5mm}
{\it Table 1:} Elastic properties of some transversely isotropic materials.
\end{center}

\section{The indentation problem}

In this section, we shall develop an approximate analytical solution to the problem of a rigid sphere of radius $R$ indenting an anisotropic half space whose Green's function is defined by equations (\ref{G3}, \ref{htheta}), including the effects of adhesion. We remark here that the more general problem involving two deformable spheres with radii $R_1, R_2$ and elastic compliance functions $h_1(\theta), h_2(\theta)$ is readily solved by substituting
\begin{equation}
\frac{1}{R}=\frac{1}{R_1}+\frac{1}{R_2}\;;\;\;\;h(\theta)=h_1(\theta)+h_2(\theta)
\end{equation}
in the following equations.

If there were no adhesion, the contact area between any two quadratic elastic bodies would be elliptical and the contact pressure distribution would have the Hertzian form 
\begin{equation}
p_H(x,y)=p_0\sqrt{1-\frac{x^2}{a^2}-\frac{y^2}{b^2}}\;,  \label{Hertz}
\end{equation}
where $a,b$ are the semi-axes of the ellipse and $p_0$ is a constant. This result applies for generally anisotropic materials and was established by Willis \cite{Willis-Hertz}. One might expect that the corresponding JKR solution involving adhesive forces could be obtained as in the original axisymmetric solution \cite{JKR} by superposing an appropriate multiple of the pressure distribution 
\begin{equation}
p_F(x,y)=\frac{1}{\sqrt{1-x^2/a^2-y^2/b^2}}\;,
\end{equation}
which can be shown to cause a uniform normal displacement over the elliptical contact area \cite{Willis-Flat}. However, this superposition leads to a stress-intensity factor at the edge of the contact area that varies around the ellipse. It follows that the contact area for the adhesive problem will not generally be strictly elliptical except in the circular limit $a=b$, though it might be anticipated that deviations from the elliptical shape would be small. Johnson and Greenwood \cite{JKR-elliptical} obtained an approximate solution for the related problem of adhesive contact of isotropic ellipsoidal bodies by assuming an elliptical contact area with the pressure distribution 
\begin{equation}
p(x,y)=\frac{B_0+B_1x^2/a^2+B_2y^2/b^2}{\sqrt{1-x^2/a^2-y^2/b^2}}\;,
\label{pxy}
\end{equation}
where $B_0,B_1,B_2$ are three constants that are chosen so as to give the correct stress-intensity factor at the ends of the major and minor axes and to satisfy the contact condition within the ellipse. They found that with this assumption, the maximum deviation from the correct stress-intensity factor was of the order of 5\% and occurred approximately midway between the pairs of points $(\pm a,0)$ and $(0,\pm b)$. In this paper, we shall apply Johnson and Greenwood's method to obtain an approximate solution for the case where quasi-eccentricity of the contact area is due to material anisotropy, rather than the indenter geometry.

\subsection{Determination of the surface displacements}

\begin{center}
\includegraphics[height=52mm]{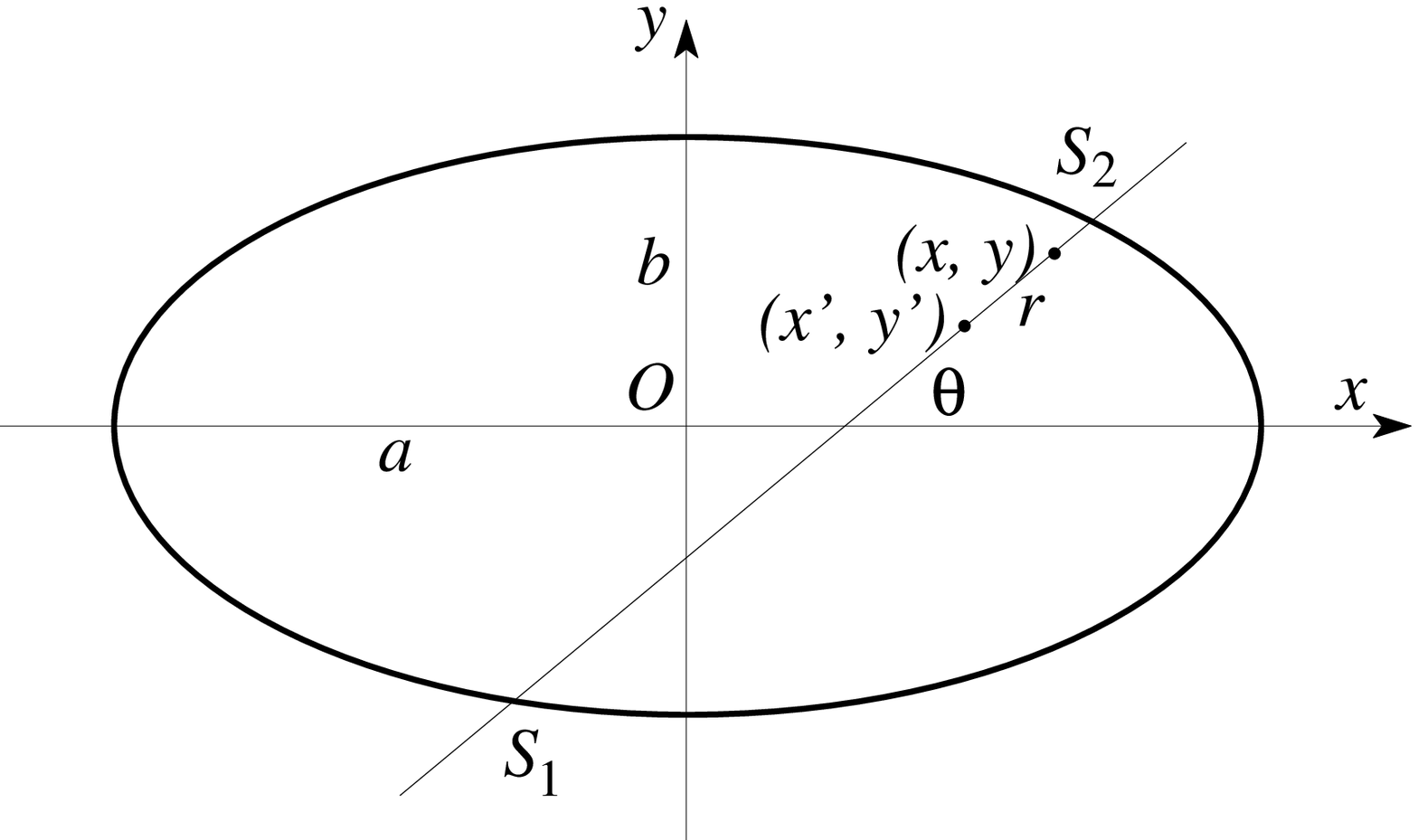}

\vspace{0.2in} \noindent \textit{Figure 1:} Geometry for field-point
integration as in equation (\ref{1}).
\end{center}

\vspace{0.2in} If the contact area is the ellipse shown in Figure 1, the Green's function (\ref{G3}) can be used to write the
inward normal displacement inside the contact area in the form
\begin{equation}
u(x,y)=\int_0^\pi\int_{S_1}^{S_2}p(x^\prime,y^\prime)h(\theta)drd\theta\;,
\label{1}
\end{equation}
where the pressure $p(x,y)$ is defined by equation (\ref{pxy}) and the polar coordinates $(r,\theta)$ are based on the field point $(x,y)$ as origin. We know that the displacements inside the ellipse must have the quadratic form 
\begin{equation}
u(x,y)=C_0+C_1x^2+C_2y^2+C_3xy  \label{uxy}
\end{equation}
\cite{Willis-Flat}, with the coefficients $C_0, C_1, C_2, C_3$ being linear functions of $B_0, B_1, B_2$ from equation (\ref{pxy}), and more general functions of $\theta$ and the semi-axes $a,b$. The exact form of these relations will be determined in the Appendix.

If the anisotropic half space is indented by a rigid sphere of radius $R$, we require 
\begin{equation}
C_1=C_2=-\frac{1}{2R}\;;\;\;\;C_3=0\;,  \label{C1C2R}
\end{equation}
which provides three equations for the five unknowns $B_0, B_1, B_2, a,b$ and an angle defining the orientation of the ellipse. Two further equations are obtained from the requirement that the energy release rate at the two points $(\pm a,0)$ and $(0, \pm b)$ be equal to the interface energy $\Delta\gamma$. Finally, if the total force $F$ applied to the indenter is prescribed, we have 
\begin{equation}
F=\int_{-a}^a\int_{-b\sqrt{1-x^2/a^2}}^{b\sqrt{1-x^2/a^2}}p(x,y)dydx=2\pi
ab\left(B_0+\frac{B_1}{3}+\frac{B_2}{3}\right)\;.  \label{F}
\end{equation}
The solution of this problem will define the dimensions of the contact area as a function of the applied force $F$, but the quantity of most interest is the pull-off force which comprises the maximum negative value of $F$.

In this paper, we shall restrict attention to the case where the material anisotropy exhibits a symmetry plane, in which case the coefficients $b_m$ in (\ref{htheta}) will be zero and the condition $C_3=0$ will be satisfied identically. However, we shall discuss possible strategies for solving the more general problem in Section \ref{unsym}.

\subsection{Stress-intensity factors}

The JKR theory demands that the energy release rate at the edge of the contact area be equal to the interface energy $\Delta\gamma$, which is equivalent to the condition that the local stress-intensity factor be given by
\begin{equation}
K_I=\sqrt{\frac{2\Delta\gamma}{\pi h(\theta)}}\;,
\end{equation}
where $\theta$ is the inclination of the local boundary of the contact ellipse. 

Applying this condition at the ends of the major axis $(\pm a,0)$ where $\theta=\pi/2$, we obtain
\begin{equation}
(B_0+B_1)=-\frac{1}{\pi}\sqrt{\frac{2\Delta\gamma}{h(\pi/2)a}}\;.
\label{KIa}
\end{equation}
A similar condition imposed at the points $(0,\pm b)$ yields 
\begin{equation}
(B_0+B_2)=-\frac{1}{\pi}\sqrt{\frac{2\Delta\gamma}{h(0)b}}\;.  \label{KIb}
\end{equation}

\subsection{Dimensionless formulation}

In view of the parametric dependence of the isotropic JKR solution \cite{JKR} , it is convenient to introduce the dimensionless variables 
\begin{equation}
\Lambda=\left(\frac{2\pi^2\Delta\gamma h_0}{R}\right)^{1/3}\;;\;\;\;\hat{b}=
\frac{b}{R\Lambda}\;;\;\;\;\beta_i=\frac{B_ih_0}{\Lambda}\;;\;\;\;i=0,1,2\;.
\end{equation}
We then obtain 
\begin{equation}
\hat{F}\equiv\frac{F}{\pi\Delta\gamma R}=\frac{4\pi^2\hat{b}^2}{\sqrt{1-e^2}}
\left(\beta_0+\frac{\beta_1}{3}+\frac{\beta_2}{3}\right)\;,  \label{Fnd}
\end{equation}
from (\ref{F}), and 
\begin{equation}
\beta_0+\beta_1=-\frac{1}{\pi^2}\sqrt{\frac{\sqrt{1-e^2}}{(1-\lambda_1)\hat{b}}}\;;\;\;\;
\beta_0+\beta_2=-\frac{1}{\pi^2}\sqrt{\frac{1}{(1+\lambda_2)\hat{b}}} \;,  \label{K2nd}
\end{equation}
from (\ref{KIa}, \ref{KIb}), where 
\begin{equation}
\lambda_1=1-\frac{h(\pi/2)}{h_0}=-\sum_{m=1}^\infty
(-1)^ma_m\;;\;\;\;\lambda_2=\frac{h(0)}{h_0}-1=\sum_{m=1}^\infty a_m\;.
\end{equation}
Also, the condition (\ref{C1C2R}) requires 
\begin{equation}
\phi_{11}\beta_1(1-e^2)+\phi_{21}\beta_2
=\phi_{12}\beta_1(1-e^2)+\phi_{22}\beta_2=-\frac{\hat{b}}{2\pi}\;,
\label{CRnd}
\end{equation}
where the functions $\phi_{ij}$ are defined in equation (\ref{phis}). Notice that the power series expressions (\ref{powerseries}) must be used for values of $e$ near zero, to avoid numerical errors.

\subsection{Solution strategy}

\label{strategy}

The eccentricity $e$ of the contact area varies with the force $F$, so a convenient strategy is to regard $e$ as an independent parameter. We then solve the two equations (\ref{CRnd}) for $\beta_1,\beta_2$ as functions of $
\hat{b}$, substitute the solution into the two equations (\ref{K2nd}), and eliminate $\beta_0$ to obtain an equation for $\hat{b}$. The parameters $\beta_0, \beta_1,\beta_2$ can then be determined and finally the force $\hat{F}$ is obtained from (\ref{Fnd}). Also, the central displacement $d=u(0,0)$, representing the indentation of the sphere, can then be obtained as 
\begin{equation}
\frac{d}{R\Lambda^2}\equiv\hat{d}=\pi\hat{b}\left(\phi_0\beta_0+\phi_{10}
\beta_1(1-e^2)+\phi_{20}\beta_2\right)\;,  \label{dhat}
\end{equation}
from (\ref{uxyA}).

\section{Results}

In the interests of simplicity, we restrict the numerical calculations to cases where the series in (\ref{htheta}) are truncated at $m=1$. It is then a trivial matter to choose an orientation for the coordinate system to make $b_1=0$, so that the only non-zero coefficient in (\ref{htheta}) is $a_1$, which can often be approximated by (\ref{a1-approx}) and which must then lie in the range $-1<a_1\leq1$, since $h(\theta)>0$ for all $\theta$.

Figure 2 shows the relation between the axis ratio $b/a=\sqrt{1-e^2}$ and the dimensionless force $\hat{F}$ for the case where $a_1=-0.5$ and hence $h(\pi/2)/h(0)=3$. When the force is large and compressive (positive), $b/a$ 
tends to a limiting value 0.697 which is also the value that would be obtained for any value of the indenting force in the absence of adhesive forces. This limit is shown by a vertical dashed line in Figure 2. The eccentricity changes only slightly in the compressive range $\hat{F}>0$, but in the tensile range the contact area 
becomes progressively more circular.

\begin{center}
\includegraphics[height=62mm]{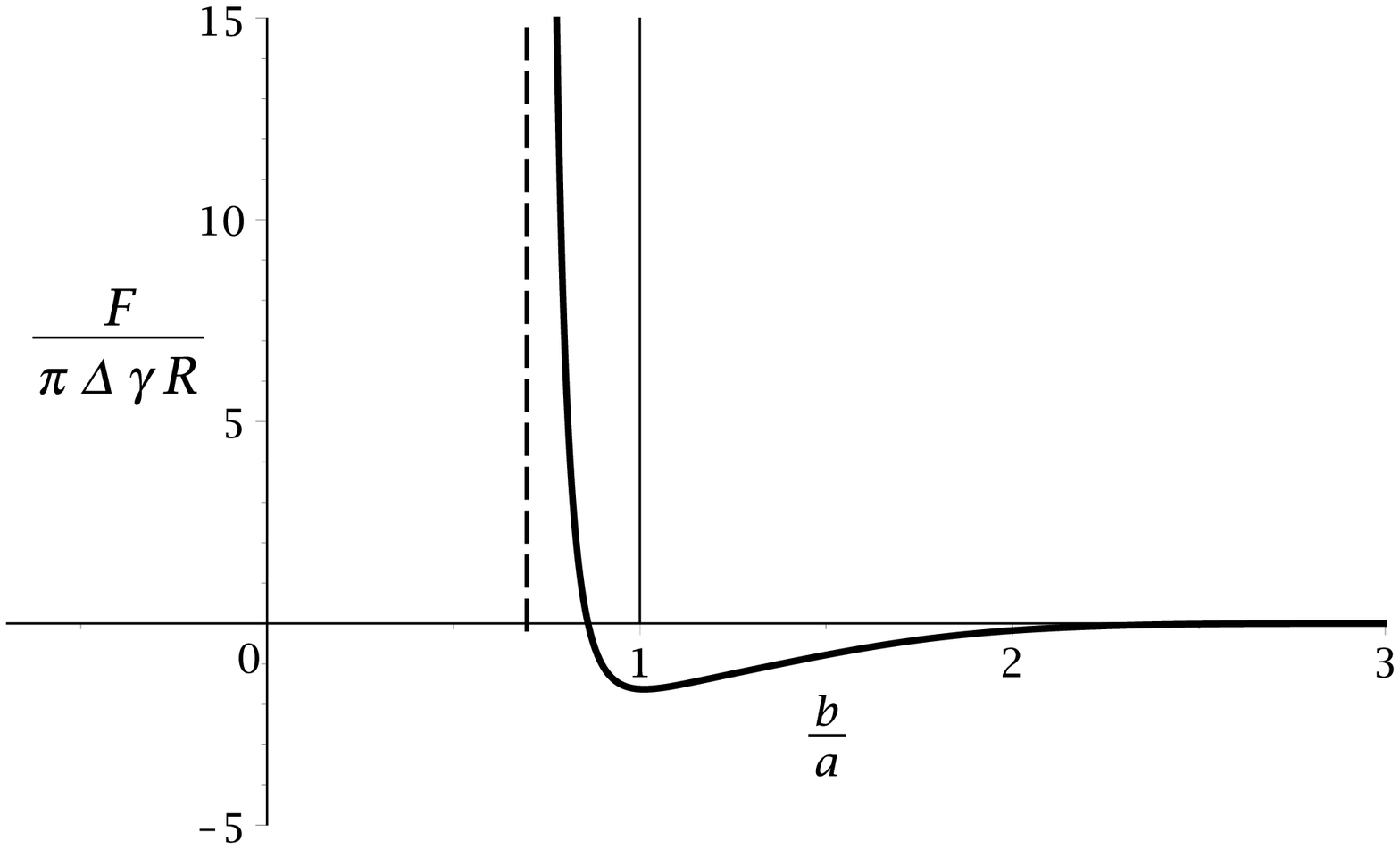}
\end{center}

\vspace{0.2in} \noindent \textit{Figure 2:} Variation of the axis ratio of the contact area with dimensionless indentation force $\hat{F}$, for $a_1=-0.5$.

\vspace{0.2in} These results are broadly similar to those of Johnson and Greenwood \cite{JKR-elliptical} for the adhesive indentation of isotropic materials by a non-spherical quadratic indenter, but one significant 
difference here is that the tensile force is still increasing when we reach the circular geometry $b/a=1$. To proceed beyond this point, we need to interchange the $x$ and $y$-axes, since the elliptic integrals are defined 
only for the case $b/a<1$. This can be done simply by changing the sign of $a_1$ and interchanging $a$ and $b$.

\begin{center}
\includegraphics[height=78mm]{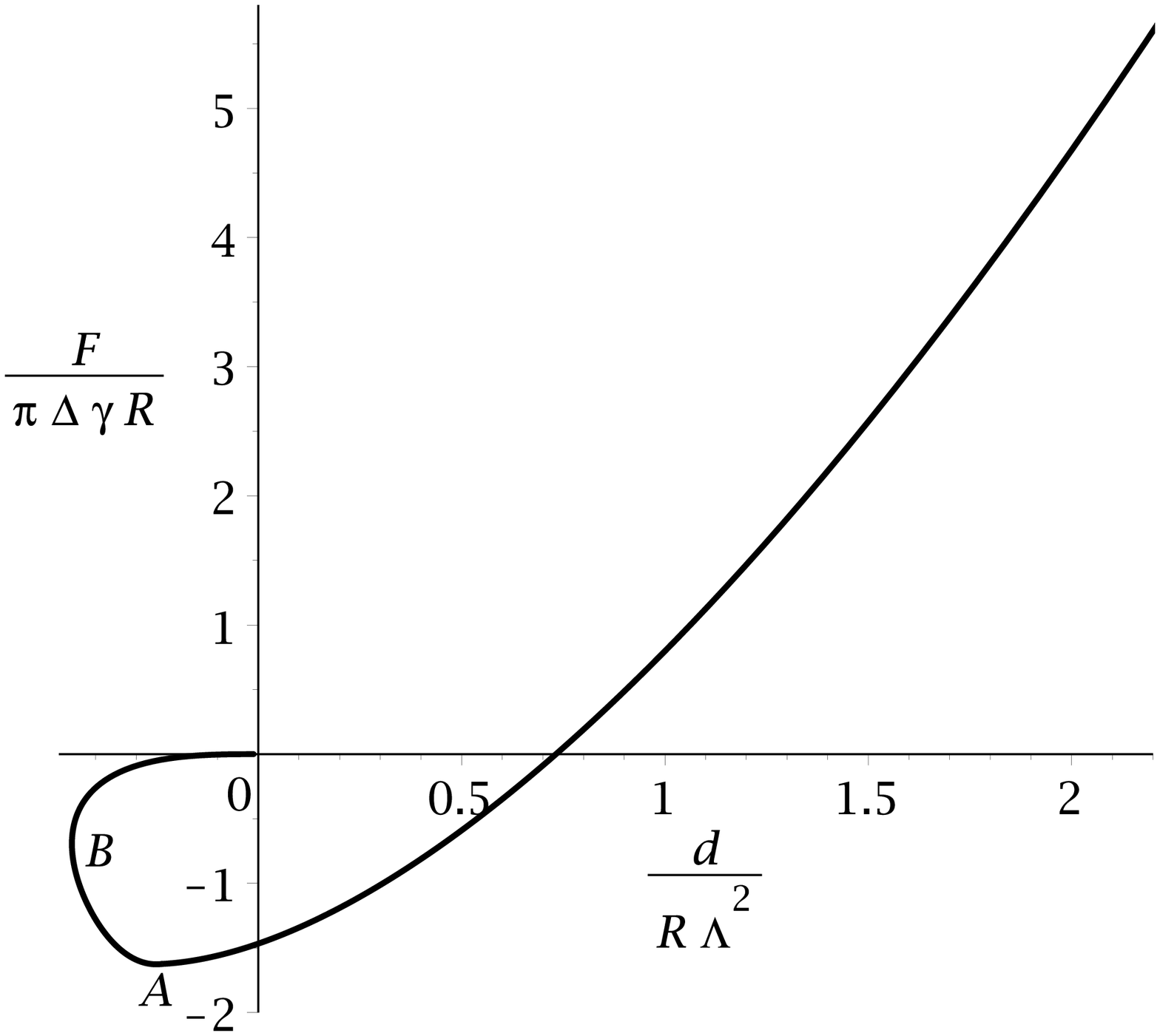}
\end{center}

\vspace{0.2in} \noindent \textit{Figure 3:} Relation between indentationforce and displacement for $a_1=-0.5$.

\vspace{0.2in}  Figure 3 shows the relation between $\hat{F}$ and the dimensionless indentation $\hat{d}$, plotted parametrically from equation (\ref{dhat}), for the case $a_1=-0.5$. The pull-off force corresponds to the point $A$ in the figure, whereas point $B$ defines the point at which pull-off would occur in a controlled-displacement experiment. As we proceed from $A$ and $B$, the contact area becomes increasingly eccentric {\it in the opposite direction to that obtained during compressive loading}.

For $a_1=-0.5$, the maximum negative value of $\hat{F}$ (the dimensionless pull-off force) occurs at $b/a=1.011$ ($e=0.15$) and is $-1.6272$. By comparison, the dimensionless force at the `circular' point $e=0$ is $\hat{F}
=-1.6244$, suggesting that a good approximation to the pull-off force can be obtained by assuming \textit{a priori} that the contact area is circular. Notice incidentally, that in this condition, the contact pressure 
distribution is \textit{not} axisymmetric, since the stress-intensity factors differ on two perpendicular axes. However, with $a_1$ the only non-zero coefficient, the circular solution then becomes exact, since both the stress-intensity factor and the directional modulus vary in the same way with $\theta$. The analysis is then greatly simplified and the force is obtained in closed form as 
\begin{equation}
\hat{F}=\frac{3(2-a_1^2)}{2\left(1+\sqrt{1-a_1^2}\right)\sqrt{1-a_1^2}}\;.
\label{circular}
\end{equation}
Also, the dimensionless radius of the contact area in this state is 
\begin{equation}
\hat{b}=\left(\frac{3(2-a_1^2)}{4\left\{\sqrt{1+a_1}+\sqrt{1-a_1}\right\}
\sqrt{1-a_1^2}}\right)^{2/3}\;.
\end{equation}
These results of course reduce to the classical JKR values $\hat{F}=1.5, \hat{b}=(3/4)^{2/3}$ in the isotropic case $a_1=0$.

\begin{center}
\includegraphics[height=62mm]{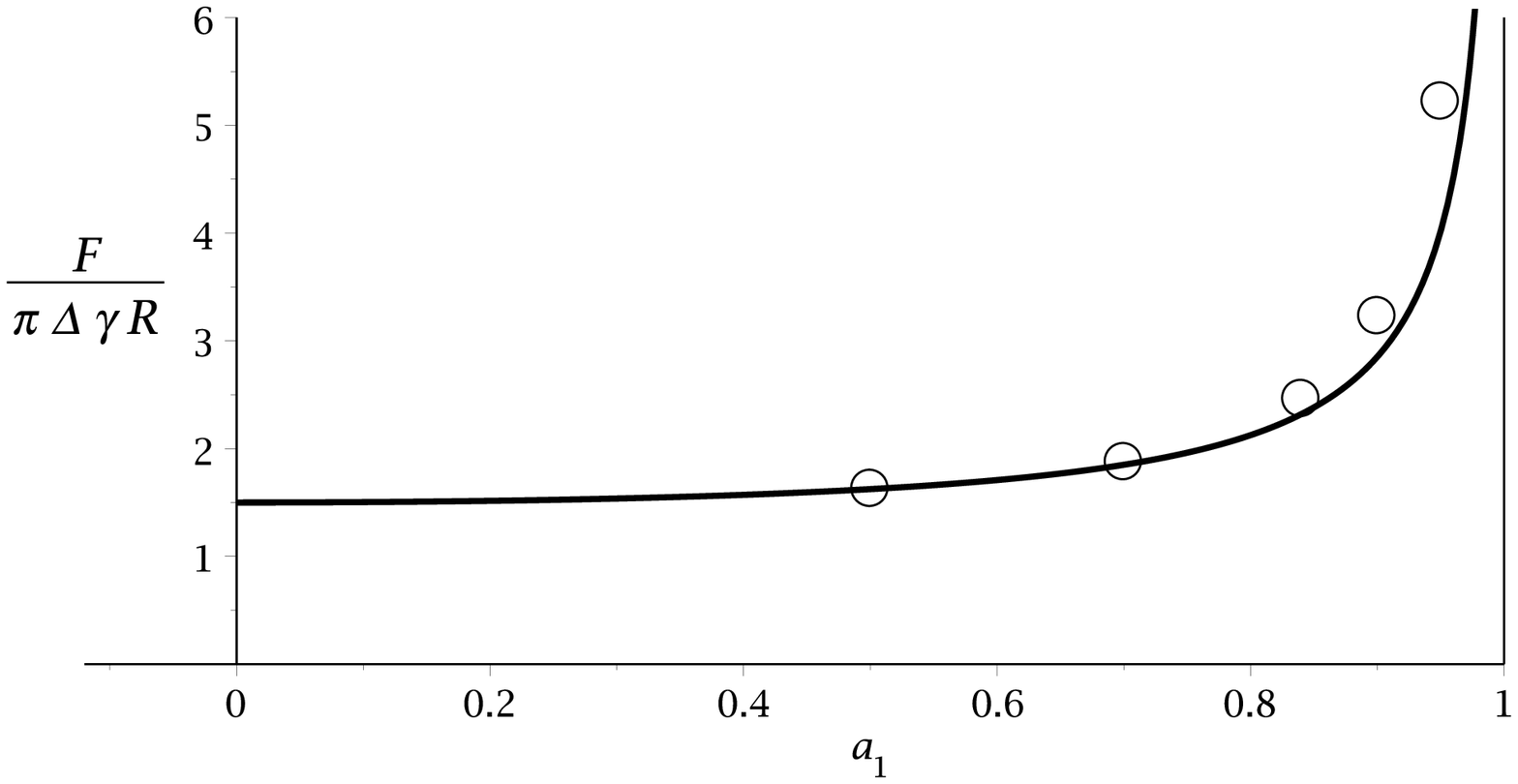}
\end{center}

\vspace{0.2in} \noindent \textit{Figure 4:} Pull-off force as a function of $a_1$ (points). The solid line represents equation (\ref{circular}), based on the approximation that the contact area at pull-off is circular.

\vspace{0.2in} The solid line in Figure 4 shows the pull-off force predicted by this `circular' approximation, whereas the points represent the more exact pull-off force found by iterating on the eccentricity until the maximum tensile force is obtained. Notice that $\hat{F}$ is independent of the sign of $a_1$. Clearly the approximation is very good for modest levels of anisotropy, but it underestimates the pull-off force for $|a_1|>0.7$.

It is notable that although the mean modulus $h_0$ has no effect on $\hat{F}$ (as can be demonstrated even for the exact solution, using similarity and dimensional arguments \cite{JRB-adhesion}), the degree of anisotropy defined through the coefficient $a_1$ [or the ratio $h(0)/h(\pi/2)$] leads to a significant increase in the pull-off force
relative the the JKR value.

\section{More general cases}\label{unsym}

The method described here can easily be extended to cases where the indenting body is ellipsoidal, or where more terms are included in the series (\ref{htheta}). However, recalling that the energy release rate condition is satisfied only at the ends of the axes of the ellipse, we must anticipate more significant errors in the satisfaction of this condition at intermediate points if these higher-order terms are significant.

If the material does not exhibit a plane of symmetry, the orientation of the ellipse is an additional unknown. The solution strategy defined in Section  \ref{strategy} can still be applied, but in general we would not expect the 
resulting pressure distribution to satisfy the condition $C_{3}=0$ in equation (\ref{uxy}). Suppose we then rotate the coordinate system by a small angle, thus redefining the Fourier coefficients in equation (\ref{htheta}). The magnitude of the coefficient $C_{3}$ for a given force $F$ will be changed, and a Newton-Raphson iterative scheme should allow the coordinate system to be rotated until the condition $C_{3}=0$ is satisfied.

\section{Conclusions}

We have developed an approximate JKR solution for the adhesive contact of anisotropic materials, by assuming an elliptical contact area and imposing the condition at the contact edges for energy release rate only at the extremes of the axes. The contact area becomes less elliptical as the compressive force is reduced and becomes elliptical in the opposite sense in the tensile r\'{e}gime. If the point force solution for the anisotropic half space is approximated by a two term Fourier series, the contact area at the pull-off force is found to be very close to circular, even though the contact pressure distribution is not axisymmetric. In this case, (i) a closed form expression can be obtained for the pull-off force and (ii) the energy release rate condition is satisfied exactly all around the contact area.
Perhaps the most remarkable conclusion is that the pull-off force is independent of the mean compliance modulus $h_0$, as in the JKR solution, but it is significantly increased by the dimensionless degree of anisotropy. 

\section{Acknowledgements}

The authors thank Mr. N. Menga from Politecnico di Bari for help in development of the calculations.

\appendix

\section{Evaluation of surface displacements}

\subsection*{The flat punch solution}

We first consider the surface displacements due to the pressure distribution 
\begin{equation}
p_0(x^{\prime},y^{\prime})=\frac{1}{\sqrt{1-x^{\prime\,2}/a^2-y^{\prime\,2}/b^2}}\;.  \label{p0}
\end{equation}
Substituting into (\ref{1}) and writing $x^{\prime}=x-r\cos\theta\;,y^{%
\prime}=y-r\sin\theta$, we obtain 
\begin{equation}
u_0(x,y)=\int_0^\pi\int_{S_1}^{S_2}\frac{h(\theta)drd\theta}{\sqrt{%
A(\theta)+B(\theta)r-C(\theta)r^2}}\;,  \label{1a}
\end{equation}
where 
\begin{equation*}
A(\theta)=1-\frac{x^2}{a^2}-\frac{y^2}{b^2}\;;\;\;\; B(\theta)=2\left(\frac{%
x\cos\theta}{a^2}+\frac{y\sin\theta}{b^2}\right) 
\end{equation*}
\begin{equation}
C(\theta)=\frac{\cos^2\theta}{a^2}+\frac{\sin^2\theta}{b^2}=\frac{%
(1-e^2\cos^2\theta)}{b^2}\;,
\end{equation}
and the eccentricity $e=\sqrt{1-b^2/a^2}$. The change of variable $t=r-B/(2C)$
yields 
\begin{equation}
u_0(x,y)=\int_0^\pi\int_{-D}^{D}\frac{h(\theta)dtd\theta}{\sqrt{C(D^2-t^2)}} 
\ssb{with}D^2=\frac{A}{C}+\frac{B^2}{4C^2}  \label{1b}
\end{equation}
and the inner integral can then be performed to give 
\begin{equation}
u_0(x,y)=\pi\int_0^\pi C^{-1/2}h(\theta)d\theta=\pi b\int_0^\pi\frac{%
h(\theta)d\theta}{(1-e^2\cos^2\theta)^{1/2}}\;.  \label{1c}
\end{equation}
Substituting for $h(\theta)$ from (\ref{htheta}), we obtain 
\begin{equation}
u_0(x,y)=\pi h_0b\sum_{m=0}^\infty a_mI_0(m,e)\ssb{where}I_0(m,e)=\int_0^\pi \frac{\cos(2m\theta)d\theta}{(1-e^2\cos^2%
\theta)^{1/2}}\;.
\end{equation}
These integrals can be evaluated in terms of complete elliptic integrals,
the first few being 
\begin{equation*}
I_{0}(0,e)=2K(e) \;;\;\;\; I_{0}(1,e)=\frac{4[K(e)-E(e)]}{e^2}-2K(e) 
\end{equation*}
\begin{equation}
I_{0}(2,e)=\frac{32[K(e)-E(e)]}{3e^4}+\frac{16[E(e)-2K(e)]}{3e^2}+2K(e)
\label{I3}
\end{equation}

These results, though exact, do not evaluate easily when $e\ll1$. In this
range it is better to use power series approximations for the elliptic
integrals, giving 
\begin{eqnarray}
I_0(0,e)&=&\pi\left(1+\frac{e^2}{4}+\frac{9e^2}{64}+\frac{25e^6}{256}+\frac{%
1225e^8}{16384}\right)+O(e^{10})  \notag \\
I_0(1,e)&=&\frac{\pi e^2}{8}\left(1+\frac{3e^2}{4}+\frac{75e^4}{128}+\frac{%
245e^6}{512}\right)+O(e^{10})  \label{powerseries} \\
I_0(2,e)&=&\frac{\pi e^4}{128}\left(3+\frac{15e^2}{4}+\frac{245e^4}{64}%
\right)+O(e^{10})  \notag
\end{eqnarray}

\subsection*{The Hertzian solution}

To determine the displacements due to the two remaining terms in (\ref{pxy}%
), it is convenient to start from the Hertzian distribution (\ref{Hertz}),
but with $p_0=1$. The same changes of variable used above yield the
displacements 
\begin{equation}
u_H(x,y)=\int_0^\pi\int_{-D}^{D}\sqrt{C(D^2-t^2)}\,h(\theta)dtd\theta\;,
\label{1d}
\end{equation}
and after performing the inner integral, the resulting function has the form
of equation (\ref{uxy}) with 
\begin{eqnarray}
C_0&=&\frac{\pi b}{2}\sum_{m=0}^\infty a_mI_0(m,e)\;;\;\;\;C_1=-\frac{\pi b}{%
2a^2}\sum_{m=0}^\infty a_mI_1(m,e)  \notag \\
C_2&=&-\frac{\pi b}{2a^2}\sum_{m=0}^\infty a_mI_2(m,e)\;;\;\;\; C_3=\frac{%
\pi b}{2a^2}\sum_{m=1}^\infty b_mI_3(m,e)\;,  \notag
\end{eqnarray}
where 
\begin{eqnarray}
I_1(m,e)&=&\int_0^\pi \frac{\cos(2m\theta)\sin^2\theta d\theta}{%
(1-e^2\cos^2\theta)^{3/2}} \;;\;\;\; I_2(m,e)=\int_0^\pi \frac{%
\cos(2m\theta)\cos^2\theta d\theta}{(1-e^2\cos^2\theta)^{3/2}}  \notag \\
I_3(m,e)&=&\int_0^\pi \frac{\sin(2m\theta)\sin(2\theta) d\theta}{%
(1-e^2\cos^2\theta)^{3/2}}  \notag \\
&=&\frac{1}{2}\left[I_1(m-1,e)+I_2(m-1,e)-I_1(m+1,e)-I_2(m+1,e)\right]\;. 
\notag
\end{eqnarray}
By differentiation and superposition it can be shown that 
\begin{equation}
I_1(m,e)=I_0(m,e)-\frac{(1-e^2)}{e}\frac{dI_0(m,e)}{de}\;;\;\;\;I_2(m,e)=%
\frac{1}{e}\frac{dI_0(m,e)}{de};\;.  \label{I1I2}
\end{equation}

\subsection*{Singular fields with quadratic displacements}

Since the pressure distribution $p_H(x,y)$ [with $p_0=1$] produces the
displacement field $u_H(x,y)$, it follows by superposition that the
distribution 
\begin{equation*}
\frac{\partial }{\partial a}\sqrt{1-\frac{x^2}{a^2}-\frac{y^2}{b^2}}=\frac{%
x^2}{a^3\sqrt{1-x^2/a^2-y^2/b^2}} 
\end{equation*}
will produce the displacement $\mbox{$\partial$} u_H(x,y)/\mbox{$\partial$} a
$. This enables us to determine the displacement due to the term $B_1$ in
equation (\ref{pxy}), and a similar procedure differentiating with respect
to $b$ yields the contribution of the term $B_2$. Using these results, and
noting for example that 
\begin{equation*}
\frac{\partial }{\partial a}I_0(m,e)=\frac{\partial }{\partial e}I_0(m,e)%
\frac{\partial e}{\partial a}=\frac{b^2}{a^3e}\frac{\partial }{\partial e}%
I_0(m,e)\;, 
\end{equation*}
the complete displacement field due to the pressure distribution (\ref{pxy})
is obtained as 
\begin{eqnarray}
u(x,y)&=&\pi h_0b\left[\phi_0B_0+\frac{\phi_{10}B_1b^2}{a^2}%
+\phi_{20}B_2+\left(\frac{\phi_{11}B_1}{a^2}+\frac{\phi_{21}B_2}{b^2}%
\right)x^2\right.  \notag \\
&&\left.+\left(\frac{\phi_{12}B_1}{a^2}+\frac{\phi_{22}B_2}{b^2}\right)y^2
+\left(\frac{\phi_{13}B_1}{a^2}+\frac{\phi_{23}B_2}{b^2}\right)xy\right]\;,
\label{uxyA}
\end{eqnarray}
where 
\begin{eqnarray}
\phi_{0}&=&\sum_{m=0}^\infty a_mI_0(m,e)\;;\;\;\;\phi_{10}=\frac{(1-e^2)}{2e}%
\sum_{m=0}^\infty a_m\frac{\partial I_0(m,e)}{\partial e}  \notag \\
\phi_{20}&=&\frac{1}{2}\sum_{m=0}^\infty a_mI_1(m,e)  \notag \\
\phi_{11}&=&\sum_{m=0}^\infty a_m\left(I_1(m,e)-\frac{(1-e^2)}{2e}\frac{%
\partial I_1(m,e)}{\partial e}\right)  \notag \\
\phi_{12}&=&\sum_{m=0}^\infty a_m\left(I_2(m,e)-\frac{(1-e^2)}{2e}\frac{%
\partial I_2(m,e)}{\partial e}\right)  \notag \\
\phi_{13}&=&-\sum_{m=1}^\infty b_m\left(I_3(m,e)-\frac{(1-e^2)}{2e}\frac{%
\partial I_3(m,e)}{\partial e}\right)  \notag \\
\phi_{21}&=&-\frac{(1-e^2)}{2}\sum_{m=0}^\infty a_m\left(I_1(m,e)-\frac{%
(1-e^2)}{e}\frac{\partial I_1(m,e)}{\partial e}\right)  \notag \\
\phi_{22}&=&-\frac{(1-e^2)}{2}\sum_{m=0}^\infty a_m\left(I_2(m,e)-\frac{%
(1-e^2)}{e}\frac{\partial I_2(m,e)}{\partial e}\right)  \notag \\
\phi_{23}&=&\frac{(1-e^2)}{2}\sum_{m=1}^\infty b_m\left(I_3(m,e)-\frac{%
(1-e^2)}{e}\frac{\partial I_3(m,e)}{\partial e}\right)  \label{phis}
\end{eqnarray}


\begin{thebibliography}{10}
\expandafter\ifx\csname url\endcsname\relax
  \def\url#1{\texttt{#1}}\fi
\expandafter\ifx\csname urlprefix\endcsname\relax\def\urlprefix{URL }\fi
\expandafter\ifx\csname href\endcsname\relax
  \def\href#1#2{#2} \def\path#1{#1}\fi

\bibitem{JKR}
K.~L. Johnson, K.~Kendall, A.~D. Roberts, Surface energy and the contact of
  elastic solids, Proc. Roy. Soc. (London) A324 (1971) 301--313.

\bibitem{MYD}
V.~M. Muller, V.~S. Yuschenko, B.~V. Derjaguin, On the influence of molecular
  forces on the deformation of an elastic sphere and its sticking to a rigid
  plane, J. Colloid Interface Sci. 77 (1980) 91--101.

\bibitem{Greenwood-adhesion}
J.~A. Greenwood, Adhesion of elastic spheres, Proc. R. Soc. Lond. A453 (1997)
  1277--1297.

\bibitem{JRB-adhesion} J. R. Barber, Similarity considerations in adhesive contact problems,  Tribology International, 67 (2013) 51--53.

\bibitem{Chaudhury}
M.~K. Chaudhury, T.~Weaver, C.~Y. Hui, E.~J. Kramer, Adhesive contact of
  cylindrical lens and a flat sheet, J. Appl. Phys. 80 (1996) 30--37.

\bibitem{Pharr1} G. M. Pharr, W. C. Oliver and F. R. Brotzen, On the generality of the relationship among contact stiffness, contact area, and elastic-modulus during indentation, Journal of Materials Research, Vol. 7 (1992), pp.613--617.

\bibitem{Pharr2} W. C. Oliver and G. M. Pharr, Measurement of hardness and elastic modulus by instrumented indentation: Advances in understanding and refinements to methodology, Journal of Materials Research, Vol. 19 (2004), pp.3--20.

\bibitem{Vlassak} J. J. Vlassak and W. D. Nix, Measuring the elastic properties
of anisotropic materials by means of indentation experiments. J.Mech.Phys
Solids Vol. 42 (1994) pp.1223--1245.   

\bibitem{Willis-Flat}
J.~R. Willis, Boussinesq problems for an anisotropic half-space, J. Mech. Phys.
  Solids 15 (1967) 331--339.

\bibitem{VlassakJRB} J. J. Vlassak, M. Ciavarella, J. R. Barber and X. Wang, The indentation modulus of elastically anisotropic materials for indenters of arbitrary shape,  J.Mech.Phys.Solids,  51 (2003), 1701--1721.

\bibitem{Gao-Pharr} Y. F. Gao and  G. M. Pharr, Multidimensional contact moduli of elastically anisotropic solids, Scripta Materialia, Vol. 57 (2007) pp.13--16.

\bibitem{Delafargue}
A.~Delafargue, F.-J. Ulm, Explicit approximations of the indentation modulus of
  elastically orthotropic solids for conical indenters, Int. J. Solids Struct.
  41 (2004) 7351--7360.

\bibitem{Freund} L. B. Freund and S. Suresh, Thin film materials, Cambridge Univ. Press, Cambridge, UK, 2003, Table 3.2.

\bibitem{Willis-Hertz}
J.~R. Willis, Hertzian contact of anisotropic bodies, J. Mech. Phys. Solids 14
  (1966) 163--176.

\bibitem{JKR-elliptical}
K.~L. Johnson, J.~A. Greenwood, An approximate {JKR} theory for elliptical
  contacts, J. Phys. D: Appl. Phys. 38 (2005) 1042--1046.
\end{thebibliography}
\end{document}